\documentclass{ws-ijmpd}
\usepackage[super]{cite}
\usepackage{xcolor}
\usepackage[most]{tcolorbox}
\usepackage[verbose,hypertexnames=false]{hyperref}
\hypersetup{colorlinks=false,allbordercolors=blue,pdfborderstyle={/S/U/W 1}}

\makeatletter
\renewcommand{\wsdraftnote}{}
\renewcommand{\catchline}[5]{\expandafter\def\expandafter\@clinebuf\expandafter
      {\@clinebuf\catchlinefont
      \noindent ITCP-2026-11\hfill QMUL-PH-26-18\par
      \noindent CCTP-2026-11\hfill\par
      }\relax\par
  }
\makeatother

\newcommand{\IR}{\mathbb{R}}
\newcommand{\OO}{\mathcal{O}}

\newcommand{\KK}{\mathcal{K}}
\newcommand{\LL}{\mathcal{L}}
\newcommand{\NN}{\mathcal{N}}
\newcommand{\disc}{\mathrm{Disc}}
\newcommand{\IT}{\mathcal{I}}
\newcommand{\beq}{\begin{equation}}
\newcommand{\eeq}{\end{equation}}
\newcommand{\bea}{\begin{eqnarray}}
\newcommand{\eea}{\end{eqnarray}}

\begin{document}

\markboth{V.~Niarchos, C.~Papageorgakis}
{Neural Networks, Dispersion Relations and the Thermal Bootstrap}

\catchline{}{}{}{}{}
%%%%%%%%%%%%%%%%%%%%%%%%%%%%%%%%%%%%%%%%%%%%%%%%%%%%%%%%%%%%%%%%%%%%

\title{Neural Networks, Dispersion Relations\\ and the Thermal Bootstrap}

\author{Vasilis Niarchos}
\address{Department of Physics, ITCP \& CCTP,\\
University of Crete, 71003 Heraklion, Greece\\
niarchos@physics.uoc.gr}

\author{Constantinos Papageorgakis}
\address{Centre for Theoretical Physics and Astronomy,\\
School of Physical and Chemical Sciences,\\
Queen Mary University of London, London E1 4NS, United Kingdom\\
c.papageorgakis@qmul.ac.uk}

\maketitle

% \begin{history}
% \received{(Day Month Year)}
% \revised{(Day Month Year)}
% \accepted{(Day Month Year)}
% \published{(Day Month Year)}
% \end{history}

\begin{abstract}
We review a framework for the conformal bootstrap that does not rely on positivity and treats the infinite tower of high-dimension OPE contributions to conformal correlators through dispersion relations and neural networks. We apply it to
scalar thermal two-point functions on $S^1\times \IR^{d-1}$. 
We discuss the stability properties of the relevant non-convex optimisation scheme and potential relations to recent discussions of smoothness properties in CFT correlators. We illustrate the numerical application of the method to Generalized Free Fields
and 4d holographic CFTs. This is a proceedings contribution to the
``Athens Workshop in Theoretical Physics: 10th Anniversary", held at the
National and Kapodistrian University of Athens on December 17-19 2025.
\end{abstract}


%%=========================================================================
\section{Introduction}\label{sec:intro}

The modern conformal bootstrap leverages symmetries and general
consistency conditions to constrain conformal field theories (CFTs)
nonperturbatively. Its most successful techniques rely on feasibility
analyses of crossing equations, recast as convex semi-definite
programming problems that exploit the positivity of OPE coefficients
in unitary theories. This approach has yielded rigorous,
high-precision data across a wide range of CFTs; see e.g.\
Refs.~\refcite{Poland:2018epd,Rychkov:2023wsd,Poland:2022qrs,Hartman:2022zik}
for recent reviews.

There are, however, physically important contexts in which positivity
is absent (e.g. finite-temperature theories, theories with defects or
boundaries, higher-point crossing equations) and in which one would
like to reconstruct full solutions of the bootstrap equations rather
than to exclude assumptions about a small set of CFT data. Such a
\emph{primal bootstrap} is harder: the equations involve continuous
families of constraints and an infinite number of operators, and
standard hard or soft truncations of the OPE introduce systematic
errors whose magnitude is typically difficult to
control~\cite{Gliozzi:2013ysa,Gliozzi:2014jsa,Gliozzi:2015qsa,Gliozzi:2016cmg,Esterlis:2016psv,Li:2017ukc,Li:2017agi,Leclair:2018trn,Kantor:2021jpz,Niarchos:2023lot}.

In Ref.~\refcite{Niarchos:2025cdg} we introduced a framework for the
primal bootstrap that circumvents both positivity and uncontrolled ad hoc
truncations. Operators above an arbitrary spin cutoff $J_*$ are
captured by a subtracted dispersion relation; the remaining low-spin sector is split, at each spin,
into a finite set of exposed CFT data and a smooth one-dimensional
\emph{tail function} that packages the entire infinite tower of
high-scaling-dimension contributions. The tails are modelled by
feed-forward neural networks and bootstrapped dynamically, alongside
the exposed data, by non-convex optimisation of the
crossing condition.\footnote{Approaching the primal formulation of the conformal bootstrap as a non-convex optimisation problem with Machine Learning algorithms was initiated in Refs.~\refcite{Kantor:2021jpz,Kantor:2021kbx} and applied to a variety of theories in Refs.~\refcite{Kantor:2022epi,Niarchos:2023lot}.} Sections~\ref{sec:kms}
through~\ref{sec:holo} review this framework for scalar thermal
two-point functions on $S^1\times \IR^{d-1}$, the setting that was
the focus of Ref.~\refcite{Niarchos:2025cdg}, and discuss its
performance on Generalized Free Fields and on 4d holographic CFTs.

A second, more conceptual discussion is taken up in
Section~\ref{sec:spectral}. In applying the framework of
Ref.~\refcite{Niarchos:2025cdg} a clean pattern emerges: once a single
correct piece of theory-specific information, an \emph{anchor}, is
supplied at an intermediate radius, the non-convex optimisation
converges accurately to the physical solution; when the
anchor is incorrect or absent, the optimisation either fails or
settles onto a smooth crossing-symmetric configuration whose
prediction is biased and broadens in spread across initialisations. This observation was the inspiration for the anchor-based approach of Refs.~\refcite{Ghosh:2026jbw,Ghosh:2026xnp}, which recovers full correlators using neural network representations. The latter approach was verified to percent-level accuracy across a broad panel of theories
and dimensions, including thermal two-point functions. In Section~\ref{sec:spectral} we comment on the potential future interplay between different neural network approaches (and other existing results in the literature) in the context of the thermal bootstrap.

Thermal CFTs and holographic thermal correlators have been the subject
of rapid recent progress, with closely related analytic and numerical
developments on their analytic structure, modularity properties and
Fourier-series representations~\cite{Arnaudo:2026der,Bajc:2025jjv,Barrat:2025twb}, on bulk-cone singularities and black-hole-interior
probes~\cite{Araya:2026shz,Ceplak:2025dds,Chakravarty:2025ncy}, on
connections to hydrodynamics and pole-skipping~\cite{Davison:2025xdj},
and on thermal correlators on line defects~\cite{Giombi:2026kdz}. The
framework reviewed here is complementary to these efforts, providing a
direct numerical bootstrap of the KMS condition at non-zero spatial
separation, where spin-dependent information about the thermal spectrum
becomes accessible.

%%=========================================================================
\section{Thermal Two-Point Functions and the KMS Condition}\label{sec:kms}

We consider $d$-dimensional CFTs on $S^1\times \IR^{d-1}$. We use
coordinates $x = (\tau, \vec x)$, where
$\tau$ parametrises the thermal circle of period $\beta$ and $\vec x$
the spatial $\IR^{d-1}$. This setup captures thermal physics at
inverse temperature $\beta$ in the infinite spatial volume limit, and
can equally be viewed as the high-temperature limit of the theory on
$S^1 \times S^{d-1}$. We focus on 2-point functions of identical
scalar operators $\phi$,
\beq\label{eq:2pt}
g(\tau, |x|) := \langle \phi(x)\phi(0) \rangle_\beta \,,
\eeq
which depend separately on $\tau$ and $|x| = \sqrt{\tau^2 + \vec x^2}$
due to the reduced $SO(d-1)$ symmetry of the background. Using this
symmetry we fix $\vec x = (\sigma,0,\ldots,0)$ and introduce
\beq\label{eq:zbarz}
z := \tau + i\sigma \,, \qquad \bar z := \tau - i \sigma \,, \qquad
z = r w \,, \qquad \bar z = r w^{-1}\,.
\eeq
In this parametrisation $w$ is a pure phase, but later it will be
continued to the full complex plane.

\subsection{The KMS condition as crossing}

The Kubo--Martin--Schwinger (KMS)    condition states that the 2-point function
is invariant under $\tau \to \beta - \tau$,\footnote{The strict KMS statement is the thermal periodicity
$g(\tau+\beta,\vec x)=g(\tau,\vec x)$; the reflection
form~\eqref{eq:kms1} follows from KMS combined with hermiticity, and
we shall refer to it as the KMS condition for brevity. See
e.g.\ Section~2.2 of Ref.~\refcite{Iliesiu:2018fao} for the strict
convention.}
\beq\label{eq:kms1}
g(\tau, r) = g(\beta - \tau, r) \,.
\eeq%
Combined with parity invariance under $\sigma \to -\sigma$, and
with $\beta = 1$, this can
be recast as a `crossing' equation
\beq\label{eq:kmscrossing}
g(z,\bar z) = g(1-z, 1-\bar z) \,.
\eeq
The importance of \eqref{eq:kmscrossing} as a nontrivial consistency
condition on thermal CFT data was first recognised by El-Showk and
Papadodimas \cite{El-Showk:2011yvt}.

\subsection{Thermal OPE}

At finite temperature, 1-point functions of primary operators are
generally non-vanishing and scale as $\beta^{-\Delta}$ for an operator
of scaling dimension $\Delta$.\footnote{Consistently, 1-point
functions of non-identity primaries vanish as $\beta \to \infty$.}
Conformal descendants contribute trivially, and the OPE of the 2-point
function admits a thermal block expansion that, setting $\beta = 1$ henceforth without loss of generality,

reads \cite{Iliesiu:2018fao}
\beq\label{eq:OPE}
g(rw, rw^{-1}) = \sum_{\OO_{\Delta,J} \in \phi \times \phi}
a_{\OO_{\Delta,J}} \, C_J^{(\nu)}\!\left(\tfrac12 (w + w^{-1})\right)
r^{\Delta - 2\Delta_\phi} \,,
\eeq
where
\beq\label{eq:aO}
a_\OO := \frac{f_{\phi\phi\OO}\, b_\OO}{c_\OO}\,
\frac{J!}{2^J (\nu)_J} \,,\qquad \nu := \frac{d-2}{2}\,,
\eeq
and $C_J^{(\nu)}(\eta)$ are Gegenbauer polynomials. Here $b_\OO$ are
the thermal 1-point coefficients, while  $f_{\phi\phi
\OO}$ and $c_\OO$ are the zero-temperature 3-point function coefficients and the 2-point normalisation of $\OO$ respectively. The expansion \eqref{eq:OPE}
converges for $r<1$. Earlier related discussions of thermal 2-point
functions and OPEs appeared in Refs.~\refcite{Katz:2014rla,Witczak-Krempa:2015pia}.

Substituting \eqref{eq:OPE} into \eqref{eq:kmscrossing} yields
\beq\label{eq:kmsOPE}
\sum_{\OO_{\Delta,J} \in \phi\times\phi} a_{\OO_{\Delta,J}}
\left[ C_J^{(\nu)}\!\left(\tfrac12(w+w^{-1})\right)
r^{\Delta - 2\Delta_\phi}
- C_J^{(\nu)}\!\left(\tfrac12(\tilde w + \tilde w^{-1})\right)
\tilde r^{\,\Delta - 2\Delta_\phi} \right] = 0 \,,
\eeq
where we parametrised $z = rw$ and $1-z = \tilde r \tilde w$. For this
equation to be valid one needs both OPEs to converge,
\beq\label{eq:convergence}
r < 1 \,, \qquad \tilde r < 1\,.
\eeq
Following the nomenclature of 4-point functions at zero temperature we
shall refer to the OPE around $r=0$ as the $s$-channel, and
to the OPE around $\tilde r = 0$ as the $t$-channel.

Equation \eqref{eq:kmsOPE} is an infinite set of sum rules for the
thermal 1-point coefficients $b_\OO$ (equivalently, $a_\OO$). In the numerical implementations of later
sections we shall use sum rules arising from the point-wise evaluation
of \eqref{eq:kmsOPE} on a grid of points in the simultaneous OPE
convergence region on the $z$-plane, although one can also act on \eqref{eq:kmsOPE} with arbitrary
linear functionals.

The goal of the thermal bootstrap is to constrain individual 1-point
coefficients assuming knowledge of the zero-temperature CFT data. As
emphasised in the introduction, this task is complicated by two
features. First, \eqref{eq:kmsOPE} involves no positivity
requirements on the $a_\OO$. Second, the solutions of the KMS
condition at a fixed spectrum of scaling dimensions are in general
non-unique, and hard truncations, even augmented by a universal
approximation for the high-dimension contributions, cannot recover
this multiplicity. An analysis of the KMS condition must
therefore incorporate the \emph{full}, infinite set of contributions
to the OPE. The next section presents the first step in this
direction.

%%=========================================================================
\section{High/Low-Spin Split and Thermal Dispersion Relations}\label{sec:split}

Converting the sum rule~\eqref{eq:kmsOPE} into a form amenable to a
direct numerical bootstrap without uncontrolled truncations requires
two ingredients. The first is a subtracted thermal dispersion relation, which
packages operators above a freely chosen spin cutoff $J_*$ into an
integrated discontinuity. The second is a split of the remaining low-spin
contributions (at each spin) into a finite set of exposed CFT data
and a set of smooth one-dimensional tail functions that carry the entire
infinite tower of high-dimension operators.

\subsection{Dispersion relation, low-spin split, and tail functions}\label{sec:dispersion}

Dispersion relations for scalar 2-point functions on $S^1\times
\IR^{d-1}$ were introduced in
Ref.~\refcite{Alday:2020eua}; Ref.~\refcite{Niarchos:2025cdg} collects a
variety of such relations, obtained either from a direct application
of Cauchy's theorem~\cite{Strat}, or from the thermal Lorentzian
inversion formula of Ref.~\refcite{Iliesiu:2018fao}.\footnote{For related applications of the thermal Lorentzian inversion formula see also Ref.~\refcite{Petkou:2018ynm}.} Here we quote, without derivation, the
subtracted dispersion relation that is central to our
approach.

The 2-point function $g(z,\bar z)$ can be expressed in terms of its
discontinuity
\beq\label{eq:disc}
\disc[g(z,\bar z)] = -i\bigg(g(z+i\epsilon,\bar z) - g(z-i\epsilon,\bar z)\bigg)
\eeq
across $z\in(-\infty,-1)\cup(1,\infty)$ as\footnote{From here on we
use the abbreviated notation $a_{\Delta,J} \equiv a_{\OO_{\Delta,J}}$.}
\bea\label{eq:disprel}
g(rw, rw^{-1}) &=& \sum_{J=0}^{J_*} \sum_\Delta a_{\Delta,J}\,
C_J^{(\nu)}\!\left(\tfrac12(w+w^{-1})\right) r^{\Delta-2\Delta_\phi}
\nonumber\\
&& + 2\left(\int_{-\infty}^{-r^{-1}} + \int_{r^{-1}}^{\infty}\right)
dw'\, \KK_{J_*}(w,w')\,
\disc\!\left[g(rw', rw'^{-1})\right]\,,
\eea
where $J_*\geq J_0$ is an arbitrary spin cutoff chosen above a value
$J_0$ fixed by the Regge behaviour of $g$, so that no additional arc
contributions are present. A comment on the variable ranges in
\eqref{eq:disprel} is in order: the integration variable $w'$ ranges
over real values with $|w'|>r^{-1}$, so that $z'=rw'\in(-\infty,-1)\cup
(1,\infty)$ probes the discontinuity of $g$ across this branch cut;
this requires the analytic continuation of $g$ in $w'$ off the unit
circle introduced in Section~\ref{sec:kms}. The external $w$, by
contrast, lies in the OPE-convergence region of the original
correlator. The $J_*$-dependent kernel takes the form
\beq\label{eq:kernel}
\KK_{J_*}(w,w') := \frac12 \KK(w,w') - w'^{-1}(w'-w'^{-1})^{2\nu}
\sum_{J=0}^{J_*} K_J\, C_J^{(\nu)}\!\left(\tfrac12(w+w^{-1})\right)
F_J(w'^{-1})\,,
\eeq
with a universal piece
\beq\label{eq:Kuniv}
\KK(w,w') := \frac{1}{2\pi w'}\frac{w'^2 - 1}{(w'-w)(w'-w^{-1})}\,,
\eeq
and
\beq\label{eq:KJFJ}
K_J := \frac{\Gamma(J+1)\Gamma(\nu)}{4\pi\,\Gamma(J+\nu)}\,,\qquad
F_J(w) = w^{J+d-2}\,{}_2F_1\!\left(J+d-2,\tfrac{d}{2}-1,J+\tfrac{d}{2},w^2\right).
\eeq
In \eqref{eq:disprel} the integrated discontinuity on the second line
captures the contribution of all operators with spin $J>J_*$, while
the first line is the truncated thermal OPE retaining operators of
spin $J\leq J_*$. The latter still involves an infinite sum over
$\Delta$ at each spin. The subtraction in \eqref{eq:kernel} of the
spin-$J\leq J_*$ piece from the universal kernel $\KK(w,w')$ causes
$\KK_{J_*}(w,w')$ to suppress the integrand of \eqref{eq:disprel}
more and more strongly away from the branch-cut endpoints $z=\pm 1$
as $J_*$ grows. This localises the integral onto the region where
the crossed-channel OPE of $\disc[g]$ is best approximated by its
leading-twist truncation and is what makes the truncation error
controllable. The parameter $J_*$ is otherwise free, and increasing it strengthens
the localisation just described.

For each spin $J\leq J_*$ we introduce an arbitrary scaling-dimension
cutoff $\Delta_*(J)$ and write
\begin{multline}\label{eq:lowspin-split}
\sum_{J=0}^{J_*}\sum_{\Delta} a_{\Delta,J}\,
C_J^{(\nu)}\!\left(\tfrac12(w+w^{-1})\right) r^{\Delta-2\Delta_\phi} =\\
= \sum_{J=0}^{J_*} C_J^{(\nu)}\!\left(\tfrac12(w+w^{-1})\right)
\left[\,\sum_{\Delta\leq \Delta_*(J)} a_{\Delta,J}\, r^{\Delta-2\Delta_\phi}
+ A_{\Delta_*(J),J}(r)\right],
\end{multline}
where the tail functions
\beq\label{eq:tails}
A_{\Delta_*(J),J}(r) := \sum_{\Delta>\Delta_*(J)} a_{\Delta,J}\,
r^{\Delta - 2\Delta_\phi}
\eeq
package the entire infinite tower of high-dimension contributions at
spin $J$. Up to the subtraction of a finite number of low-$\Delta$
terms, the $A_{\Delta_*(J),J}(r)$ are simply the Gegenbauer
projections of $g$ onto spin $J$, and as such are always well-defined
and convergent as single-variable functions of the radial coordinate.

Combining \eqref{eq:disprel} and \eqref{eq:lowspin-split} we arrive at
the master representation:
\bea\label{eq:master}
g(rw,rw^{-1}) &=& \sum_{J=0}^{J_*}\sum_{\Delta\leq\Delta_*(J)} a_{\Delta,J}\,
C_J^{(\nu)}\!\left(\tfrac12(w+w^{-1})\right) r^{\Delta-2\Delta_\phi}
\nonumber\\
&& + \sum_{J=0}^{J_*} A_{\Delta_*(J),J}(r)\,
C_J^{(\nu)}\!\left(\tfrac12(w+w^{-1})\right)
\\
&& + 2\left(\int_{-\infty}^{-r^{-1}} + \int_{r^{-1}}^\infty\right)
dw'\, \KK_{J_*}(w,w')\, \disc\!\left[g(rw', rw'^{-1})\right]\,.
\nonumber
\eea
\noindent
This is an \emph{exact} relation with freely tunable cutoffs
$(J_*,\Delta_*(J))$. A finite number of exposed CFT data enter
explicitly in the first line, the full set of high-dimension
contributions at $J\leq J_*$ is compactly captured by a finite number
of one-dimensional tail functions in the second, and all the
remaining contributions from $J>J_*$ sit inside the integrated
discontinuity on the third.

\subsection{Approximate KMS condition}\label{sec:approxKMS}

Substituting~\eqref{eq:master} into the crossing form of the KMS
condition~\eqref{eq:kmscrossing} yields an exact sum rule that
equates the difference of two copies of~\eqref{eq:master}, one in
$(r,w)$ and one in $(\tilde r,\tilde w)$ with $1-z=\tilde r\tilde w$,
to zero. Two final approximations render this sum rule amenable to
numerical optimisation. First, the integrated discontinuity in the
third line of~\eqref{eq:master} is approximated by truncating the
crossed-channel OPE of $\disc[g]$ to a finite subset of leading-twist
operators, a procedure that has proven accurate in a variety of
related contexts~\cite{Alday:2017vkk,Alday:2017zzv,Lemos:2017vnx,Liu:2018jhs,Caron-Huot:2018kta,Cardona:2018qrt,Li:2019dix,Albayrak:2019gnz,Carmi:2019cub,Alday:2019clp,Iliesiu:2018fao,Lemos:2021azv}. Second, since the crossed-channel OPE converges only
for $r,\tilde r <1$, the integration range of the discontinuity is
restricted to the subregion of common $s$- and $t$-channel
convergence, $w'\in(-2r^{-1},-r^{-1})\cup(r^{-1},2r^{-1})$. Both
systematic errors are suppressed by the kernel $\KK_{J_*}$ away from
$z=\pm 1$ and decrease with increasing $J_*$, so that a single
parameter controls the quality of the approximation. Technical
details appear in Ref.~\refcite{Niarchos:2025cdg}. Representing the
tail functions by Multi-Branch Multi-Layer-Perceptron neural networks
(described in Section~\ref{sec:nn}) with parameters
$\boldsymbol{\theta}$ and denoting the resulting approximate
integrated discontinuity by $\IT_\disc^{(\mathrm{approx})}[J_*,\{a\};
rw,rw^{-1}]$, where $\{a\}$ are the exposed coefficients retained in
the crossed channel, one arrives at the approximate KMS condition:
\begin{align}\label{eq:kmsapprox}
0 &= \sum_{J=0}^{J_*}\sum_{\Delta\leq \Delta_*(J)} a_{\Delta,J}\bigg[
C_J^{(\nu)}\!\left(\tfrac12(w+w^{-1})\right) r^{\Delta-2\Delta_\phi}
- C_J^{(\nu)}\!\left(\tfrac12(\tilde w+\tilde w^{-1})\right)
\tilde r^{\Delta-2\Delta_\phi}\bigg]
\nonumber\cr
& + \sum_{J=0}^{J_*}\bigg[A_{\Delta_*(J),J;\boldsymbol{\theta}}(r)\,
C_J^{(\nu)}\!\left(\tfrac12(w+w^{-1})\right)
- A_{\Delta_*(J),J;\boldsymbol{\theta}}(\tilde r)\,
C_J^{(\nu)}\!\left(\tfrac12(\tilde w+\tilde w^{-1})\right)\bigg]
\\
&\qquad + \IT_\disc^{(\mathrm{approx})}[J_*,\{a\};rw,rw^{-1}]
- \IT_\disc^{(\mathrm{approx})}[J_*,\{a\};
\tilde r\tilde w,\tilde r\tilde w^{-1}]\,.
\end{align}
\noindent
The coefficients $\{a\}$ appearing in
$\IT_\disc^{(\mathrm{approx})}$ may or may not overlap with the
exposed ones on the first line. If an operator $\OO_{\Delta,J}$ with
$J\leq J_*$ does not appear explicitly on the first line, its
contribution is necessarily absorbed into the tail function
$A_{\Delta_*(J),J;\boldsymbol{\theta}}(r)$.

Several features of \eqref{eq:kmsapprox} are worth highlighting.
\begin{romanlist}[(iv)]
\item The parameter $J_*$ is tunable but not cost-free: small $J_*$
minimises the number of tail functions and is therefore numerically
light, but degrades the accuracy of $\IT_\disc^{(\mathrm{approx})}$;
large $J_*$ has the opposite behaviour. An intermediate value (in
practice $J_*\sim 4\text{-}10$) is usually optimal.
\item The dimension cutoffs $\Delta_*(J)$ do \emph{not} affect the
accuracy of \eqref{eq:kmsapprox}. They only control how many terms
are exposed explicitly on the first line. When an operator is
included in this truncated OPE we say that it is \emph{exposed}.
\item By bootstrapping the tail functions dynamically,
\eqref{eq:kmsapprox} retains the full infinite spectrum of
high-$\Delta$ contributions at $J\leq J_*$. Any truncation scheme
with a finite number of unknown numerical parameters, by contrast,
reduces the KMS condition to a linear regression problem with a
unique solution, which clashes with the known multi-solution
structure of the KMS condition (most notably in holographic CFTs,
where continuous families of solutions exist
\cite{Niarchos:2025cdg}). The non-convex character restored by the
tails is precisely what makes the detection of multiple solutions
possible.
\item In guided searches, the discontinuity term acts as a source for
the exposed $a_{\Delta,J}$ and for the tail functions: it injects
theory-specific information while the remaining unknowns adjust to
satisfy the KMS condition.
\end{romanlist}
We will return to the implications of points (iii) and (iv) in the
holographic analysis of Section~\ref{sec:holo}.

%%=========================================================================
\section{Neural Networks and Loss Functions}\label{sec:nn}

To cast \eqref{eq:kmsapprox} as a numerical bootstrap problem we need
two further ingredients: a flexible parametric family for the tail
functions $A_{\Delta_*(J),J}(r)$, and a non-negative loss functional
whose minima correspond to approximate solutions of the KMS condition.
In this section we review the choices made in Ref.\
\refcite{Niarchos:2025cdg} for both.

\subsection{Multi-Branch MLP representation of the tails}\label{sec:MBMLP}

We represent each tail function by a feedforward Multi-Layer
Perceptron (MLP) neural network (NN). The tail functions $A_{\Delta_*(J),J}(r)$ for
$J=0,2,\ldots,J_*$ depend on a single real variable.\footnote{We assume even
$J_*$ throughout since for identical scalars the OPE contains only even spins.} Rather than use a separate MLP per tail, we adopt
a sparse \emph{Multi-Branch} architecture in which a shared input
layer feeds $\tfrac{J_*}{2}+1$ parallel subnets, each with two hidden
layers and a single output. The shared input stage encodes the
radial coordinate once across all spins while each branch specialises
to its tail; this is empirically much more efficient than a fully
connected MLP with multiple output channels at equal parameter count
\cite{Niarchos:2025cdg}.

In practice the Multi-Branch MLPs used carry
$\mathcal{O}(10^4)$ optimisable parameters, collectively denoted by
$\boldsymbol{\theta}$, with output
$A_{\Delta_*(J),J;\boldsymbol{\theta}}(r)$ on branch $J$. The input
is taken to be $r^2$ for the GFF benchmark of Section~\ref{sec:gff}
(exploiting the $r\to -r$ symmetry of the GFF tails) and $r$ for the
holographic CFTs of Section~\ref{sec:holo}, where the
energy-momentum sector breaks that symmetry. A schematic of the
network is shown in Fig.~\ref{fig:NN}; full hyperparameter choices
are given in Appendix A of Ref.~\refcite{Niarchos:2025cdg}.

\begin{figure}[t]
\centerline{\includegraphics[width=0.85\linewidth]{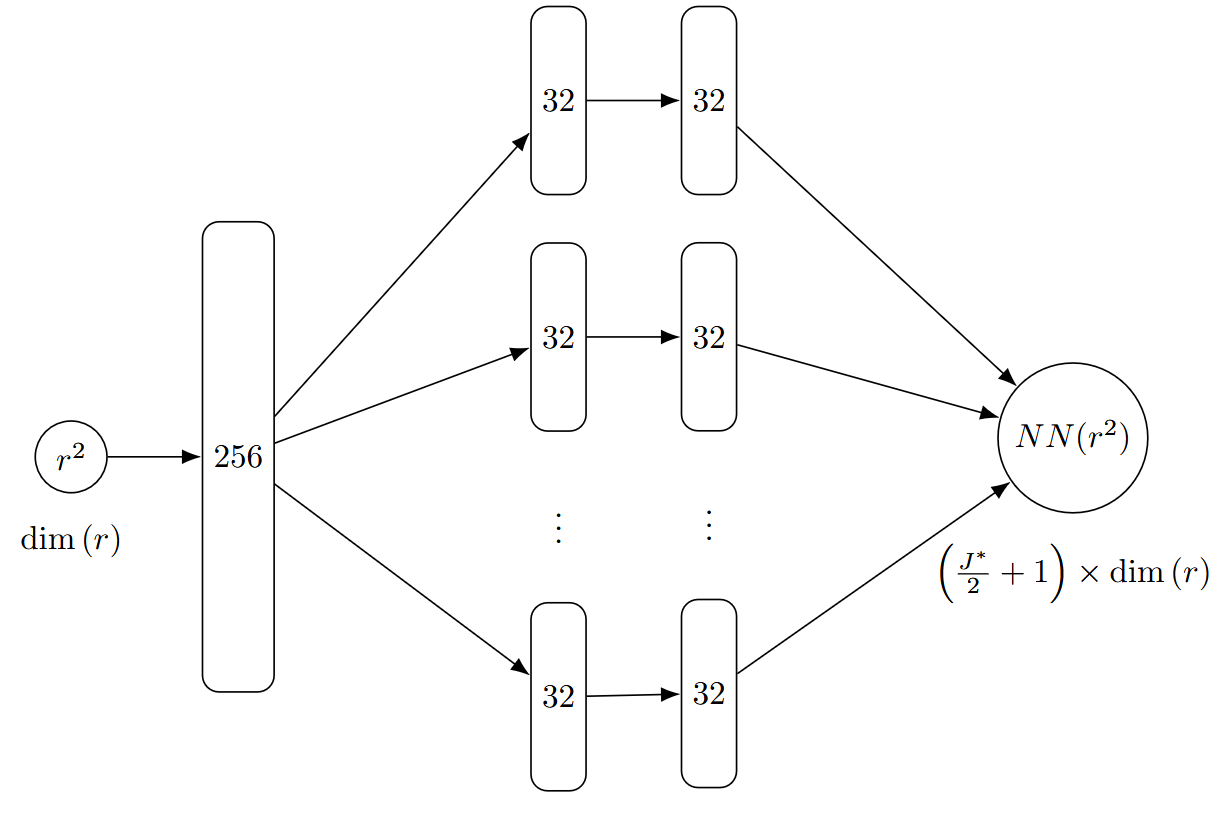}}
\vspace*{8pt}
\caption[Schematic of the Multi-Branch MLP used to parametrise the
tail functions]{Schematic of the Multi-Branch MLP used to
parametrise the tail functions in the GFF benchmark of
Section~\ref{sec:gff}. A shared backbone of width $n_b$ processes
the input $r^2$ (for GFF; $r$ in the holographic case of
Section~\ref{sec:holo}) and feeds $(J_*/2)+1$ parallel subnets of
width $n_s$, each specialising in one tail
$A_{\Delta_*(J),J;\boldsymbol{\theta}}(r)$. Tanh activation functions are used
throughout. Reproduced from Ref.~\refcite{Niarchos:2025cdg}.}%
\label{fig:NN}
\end{figure}

\subsection{KMS as a non-convex optimisation problem}\label{sec:optim}

To turn \eqref{eq:kmsapprox} into a numerical problem we discretise
$(z,\bar z)=(rw,rw^{-1})$ on a grid of finite points inside the common
$s$- and $t$-channel convergence region, and evaluate the left-hand
side pointwise to form a finite-dimensional real vector
$\vec F(\vec a,\boldsymbol{\theta})$ that depends algebraically on the
exposed coefficients $\vec a=\{a_{\Delta,J}\}$ and on the NN
parameters $\boldsymbol{\theta}$. Given a non-negative loss
$\LL(\vec a,\boldsymbol{\theta})$, one then solves the optimisation
problem
\beq\label{eq:argmin}
(\vec a_*,\boldsymbol{\theta}_*) =
\arg\min\, \LL(\vec a, \boldsymbol{\theta})\,.
\eeq
In all examples below we use \texttt{Adam}, a stochastic-gradient
optimiser with adaptive learning rates and momentum, and work on a
uniform grid of $243$ points covering $|z|<0.95$, $|1-z|<0.95$ and
slightly displaced from the real axis.\footnote{This is because $C_J^{(\nu)}(1)=J+1$ in $d=4$ collapses different-spin operators of
the same scaling dimension to identical functional forms in $\tau$,
making spin-resolved tail functions hard to disentangle.} The exact
grid is depicted in Fig.~1 of Ref.~\refcite{Niarchos:2025cdg}.

Three loss functions are used in
this work, with each targeting a different setting and exploiting a different
aspect of the structure of the KMS vector $\vec F$. The simplest, most generic option is the \emph{mean absolute loss},
\beq\label{eq:Labs}
\LL_{\overline{\mathrm{abs}}}(\vec F) := \frac{1}{N}\sum_{i=1}^N |F_i|\,,
\eeq
where $N=243$ is the number of grid points and $F_i$ are the
components of $\vec F$. This loss can be used in any context and
with any set of optimisable parameters; its main limitation is that a
good minimum value cannot be determined \emph{a priori}, making it
hard to assess quantitatively how close one is to a genuine solution
of the KMS condition.

When no exposed coefficients are optimised, $\vec F$ decomposes as
$\vec F = \vec g_{\boldsymbol{\theta}} - \vec h$ with $\vec h$ a known
vector, and we use instead the \emph{dot-product loss}
\beq\label{eq:Ldot0}
\LL_{\mathrm{dot}(0)}\;:=\;1 - \frac{|\vec g_{\boldsymbol{\theta}}\cdot \vec h|}{|\vec g_{\boldsymbol{\theta}}|\,|\vec h|}
\;+\; \bigg|1 - \frac{\vec g_{\boldsymbol{\theta}}\cdot \vec h}{|\vec h|^2}\bigg|\,.
\eeq
The first term penalises any component of $\vec g_{\boldsymbol{\theta}}$
orthogonal to $\vec h$, the second any mismatch in magnitude along
$\vec h$; both vanish at the solution. Empirically this yields
lower training losses than $\LL_{\overline{\mathrm{abs}}}$ in this
setting.

When a single exposed coefficient $a$ is optimised alongside the
tails, $\vec F = a\vec f + \vec g_{\boldsymbol{\theta}} - \vec h$ with
$\vec f$ the known crossed thermal block of the exposed operator. The
variant
\beq\label{eq:Ldot1}
\LL_{\mathrm{dot}(1)}\;:=\;1 - \frac{|(\vec g_{\boldsymbol{\theta}}-\vec h)\cdot \vec f|}
{|\vec g_{\boldsymbol{\theta}}-\vec h|\,|\vec f|}
\eeq
is minimised over $\boldsymbol{\theta}$ alone; the exposed
coefficient is recovered in closed form by the final linear projection
\beq\label{eq:aclosed}
a = -\frac{(\vec g_{\boldsymbol{\theta}_*}-\vec h)\cdot \vec f}{|\vec f|^2}\,.
\eeq
The extension to several exposed coefficients is straightforward;
throughout Sections~\ref{sec:gff}-\ref{sec:holo} we use at most one
exposed coefficient per search.

%%=========================================================================
\section{Benchmark: Generalized Free Fields}\label{sec:gff}

Generalized Free Fields (GFFs) provide an ideal first testing ground
for the framework of Sections~\ref{sec:split}-\ref{sec:nn}. In the
GFF CFT of a scalar primary $\phi$ of scaling dimension $\Delta_\phi$
in $d\geq 2$ dimensions, the thermal 2-point function admits the
closed-form image-sum representation
\beq\label{eq:GFFimages}
g(z,\bar z) = \sum_{m=-\infty}^{\infty}
\frac{1}{\big[(m-z)(m-\bar z)\big]^{\Delta_\phi}}\,,
\eeq
and its conformal block expansion receives contributions only from
the identity and the double-twist operators
$[\phi\phi]_{n,J}$, of scaling dimension
$\Delta_{n,J} = 2\Delta_\phi + 2n + J$ and even spin $J=2\ell$. The
thermal OPE coefficients take the analytic form
\beq\label{eq:aGFF}
a_{n,J} = 2\zeta(2\Delta_\phi + 2n + J)\,
\frac{(J+\nu)(\Delta_\phi)_{J+n}(\Delta_\phi - \nu)_n}{n!\,(\nu)_{J+n+1}}\,,
\eeq
where $(a)_n = \Gamma(a+n)/\Gamma(a)$ is the Pochhammer symbol.

A further simplification in the GFF case is that the crossed-channel
OPE of $\disc[g]$ receives contributions \emph{only from the identity
operator}. As a consequence, the approximate integrated discontinuity
$\IT_\disc^{(\mathrm{approx})}[J_*;\cdot]$ reduces to a fixed function
of the spacetime coordinates independent of all unknown thermal
1-point data. Concretely, for non-integer $\Delta_\phi$,\footnote{For integer
$\Delta_\phi$ the branch cuts degenerate to poles and the
discontinuity receives $\delta$-function contributions. In theories where
multi-stress-tensor and double-twist operators are degenerate (e.g.\
holographic CFTs) the OPE additionally
contains $\log(z\bar z)$ terms.}
\begin{multline}\label{eq:GFFdisc}
\IT_\disc^{(\mathrm{approx})}[J_*;rw,rw^{-1}] = \\
= 4\sin(\pi\Delta_\phi)\bigg[
\int_{r^{-1}}^{2r^{-1}} dw'\, \KK_{J_*}(w,w')\,
(rw'-1)^{-\Delta_\phi}(1-rw'^{-1})^{-\Delta_\phi} \\
- \int_{-2r^{-1}}^{-r^{-1}} dw'\, \KK_{J_*}(w,w')\,
(-rw'-1)^{-\Delta_\phi}(1+rw'^{-1})^{-\Delta_\phi}\bigg]\,.
\end{multline}
This
simplification allows us to isolate the performance of the method
from the complications of a theory-dependent discontinuity, which we
address in Section~\ref{sec:holo}.

In what follows we report tests in $d=4$ with $\Delta_\phi=1.68$.
Analogous results were obtained in Ref.~\refcite{Niarchos:2025cdg} for other
values of $\Delta_\phi$ but are not reproduced here. All optimisation
uses the 243-point $(z,\bar z)$ grid discussed in
Section~\ref{sec:optim}.

\subsection{Consistency of the approximate KMS condition}\label{sec:gff-approxKMS}

As a preliminary sanity check, we substitute the exact analytic
values of the tail functions \eqref{eq:aGFF} into the approximate KMS
condition \eqref{eq:kmsapprox} and evaluate the loss functions of
Section~\ref{sec:nn}. Low loss values at the analytic solution mean
that the scheme is based on the correct equations and that the
systematic error from the crossed-channel approximation of the
discontinuity is under control. The resulting loss values for even
$J_*=0,2,4,6,8$ with only the identity exposed and all double-twist
data absorbed into the tails are displayed in the upper block of
Table~\ref{tab:gff_analytic}. Both
$\LL_{\overline{\mathrm{abs}}}$ and $\LL_{\mathrm{dot}(0)}$ decrease
monotonically with $J_*$, reaching $\sim 10^{-4}$ and $\sim 10^{-10}$
respectively at $J_*=8$.

Before discussing tests with an optimisable exposed coefficient, we
note a peculiar analytic degeneracy of the KMS condition in $d=4$.
The contribution of a double-twist operator $[\phi\phi]_{n,2\ell}$ to
the crossed thermal block expansion takes the form
\beq\label{eq:nJdiff}
a_{n,2\ell}\bigg[r^{2n+J}C^{(\nu)}_J\!\left(\tfrac12(w+w^{-1})\right)
- \tilde r^{2n+J} C^{(\nu)}_J\!\left(\tfrac12(\tilde w+\tilde w^{-1})\right)\bigg]\,,
\eeq
and for $(n,2\ell)\in\{(1,0),(0,2)\}$ we have the identity\footnote{In~\eqref{eq:1002id} the $\cdot$ in each Gegenbauer abbreviates the argument $\tfrac12(w+w^{-1})$ for the $r^2$ terms and $\tfrac12(\tilde w+\tilde w^{-1})$ for the $\tilde r^{\,2}$ terms, as in~\eqref{eq:nJdiff}.}
\beq\label{eq:1002id}
r^2 C^{(\nu)}_0(\cdot) - \tilde r^{\,2} C^{(\nu)}_0(\cdot) = \frac{1}{\nu(2\nu+1)}\Big[
r^2 C^{(\nu)}_2(\cdot) - \tilde r^{\,2} C^{(\nu)}_2(\cdot)\Big]
= r(w+w^{-1})-1\,,
\eeq
so that the leading scalar and the leading spin-2 double-twists enter
the KMS condition only through the combination
\beq\label{eq:a10p3a02}
a_{\Delta=2\Delta_\phi+2} := a_{1,0} + \nu(2\nu+1)\,a_{0,2}\;
\overset{d=4}{=}\;a_{1,0} + 3\,a_{0,2}\,.
\eeq
This is precisely the combination of spin-dependent coefficients that
appears in the zero-spatial-separation analysis of
Ref.~\refcite{Marchetto:2023xap}. Exposing $[\phi\phi]_{1,0}$ and
$[\phi\phi]_{0,2}$ together, and evaluating
$\LL_{\mathrm{dot}(1)}$ at the analytic tails, yields the lower block
of Table~\ref{tab:gff_analytic}. The loss again decreases by several
orders of magnitude as $J_*$ grows, and the combination
\eqref{eq:a10p3a02} extracted via the closed-form prescription
\eqref{eq:aclosed} converges monotonically towards the exact analytic
value $15.06013$.

\begin{table}[t]
\tbl{Loss values on the \emph{analytic} GFF solution in $d=4$,
$\Delta_\phi=1.68$, for various $J_*$. Upper block: identity
exposed; lower block: identity and the combination $a_{1,0}+3a_{0,2}$
exposed. In the latter case, the last column reports the value of
$a_{1,0}+3a_{0,2}$ extracted from Eq.~\eqref{eq:aclosed}.\label{tab:gff_analytic}}
{\tabcolsep14pt\begin{tabular}{@{}cccc@{}}
\toprule
$J_*$ & $\LL_{\overline{\mathrm{abs}}}$ & $\LL_{\mathrm{dot}(0)}$ & \\
\colrule
0 & $0.1177$ & $0.0125$ & \\
2 & $0.0300$ & $1.3\times 10^{-5}$ & \\
4 & $0.0067$ & $1.6\times 10^{-7}$ & \\
6 & $0.0013$ & $3.3\times 10^{-9}$ & \\
8 & $0.0002$ & $0.8\times 10^{-10}$ & \\
\colrule
$J_*$ & $\LL_{\mathrm{dot}(1)}$ & & $a_{1,0}+3 a_{0,2}$ \\
\colrule
2 & $9.5\times 10^{-6}$ & & $15.16582$ \\
4 & $1.4\times 10^{-6}$ & & $15.07614$ \\
6 & $7.8\times 10^{-8}$ & & $15.06252$ \\
8 & $3.5\times 10^{-9}$ & & $15.06049$ \\
\colrule
\multicolumn{3}{r}{Exact value:} & $15.06013$ \\
\botrule
\end{tabular}}
\end{table}

\subsection{Tail bootstrap}\label{sec:tail-bootstrap}

We now \emph{bootstrap} the GFF correlator: no analytic thermal data
are inserted and the goal is to recover the GFF tails
$A_0, A_2, A_4, A_6$ from the approximate KMS condition alone. We
set $J_*=6$, expose only the identity, and initially use the loss
$\LL_{\mathrm{dot}(0)}$.

Restricting the training grid to $|z|, |1-z|<0.95$ (rather than the
full OPE-convergence domain $|z|, |1-z|<1$) leaves a rich landscape
of spurious low-loss minima with tail asymptotics inconsistent with
the KMS condition near $r=1$; we suppress these by supplementing the
loss with information about the universal $r\to 1^-$ behaviour of
the tail functions, derived in Ref.~\refcite{Niarchos:2025cdg}.

Collecting statistics from the 10 lowest-loss configurations of
$1\mathrm{K}$ independent runs of $50\mathrm{K}$ epochs with
$\LL_{\mathrm{dot}(0)}$, we obtain a mean loss
$3.54\times 10^{-9} \pm 6.4\times 10^{-10}$, comparable to the loss
$3.35\times 10^{-9}$ achieved by the analytic GFF solution. The
higher-spin tails $A_4, A_6$ are recovered accurately with small
variance. The leading tails $A_0, A_2$ exhibit significant
run-to-run variation and mean curves that deviate from the analytic
ones. Their combined contribution to the thermal OPE at $w=1$ is
however better reproduced, consistent with the expectation that our
setup has no difficulty with the spin-summed data
$\sum_J A_J(r)\, C_J^{(\nu)}(1)$ accessed by zero-spatial-separation analyses
\cite{Marchetto:2023xap,Buric:2025anb,Barrat:2025nvu}. We also
verified that the low-loss configurations satisfy the KMS condition
well on a validation grid near $z=\tfrac12$, but that it deteriorates outside the training region as a
result of the missing boundary information.

\subsection{Recovering
\texorpdfstring{$a_{1,0}+3 a_{0,2}$}{a\_\{1,0\}+3 a\_\{0,2\}}}\label{sec:tail-aDJ}

To extract the exposed combination \eqref{eq:a10p3a02} dynamically
we use the loss $\LL_{\mathrm{dot}(1)}$ and apply the closed-form
projection \eqref{eq:aclosed} at the end of the optimisation. The
universal $r\to 1^-$ asymptotics referred to above already constrains
the tails near $r=1$; we now ask what is gained by also supplying known
information about the tails at intermediate radii. We refer to such finite-$r$ constraints, which supply known values of the tail functions, as \emph{anchors} (following the terminology of Refs.~\refcite{Ghosh:2026jbw,Ghosh:2026xnp}).  Collecting the four tail
values at a chosen radius into the vector $\vec A(r) := \big(A_0(r),
A_2(r), A_4(r), A_6(r)\big)$, three anchor
choices are illustrative:\footnote{For comparative purposes, the analytic value of this datum is $a_{1,0}+3 a_{0,2}= 15.06013$ and the error we report represents ensemble
statistics from the 10 lowest-loss configurations of $1\mathrm{K}$
runs.}
\begin{itemlist}
\item An asymptotic anchor only (no finite-$r$ input) gives
$a_{1,0}+3a_{0,2} = 13.29\pm 2.82$, with the broad spread tied to the
$A_0, A_2$ ambiguity.
\item A correct analytic anchor at intermediate radius
($A_J(0.7)$ set to the GFF values) gives $15.0647\pm 0.0291$, in
excellent agreement with the analytic answer.
\item An incorrect anchor at $r=0.7$ (a random rotation of $\vec A_{\rm GFF}(0.7)$) gives $28.78\pm 4.08$, grossly wrong despite
comparable training loss.
\end{itemlist}

\begin{figure}[t]
\centerline{\includegraphics[width=\linewidth]{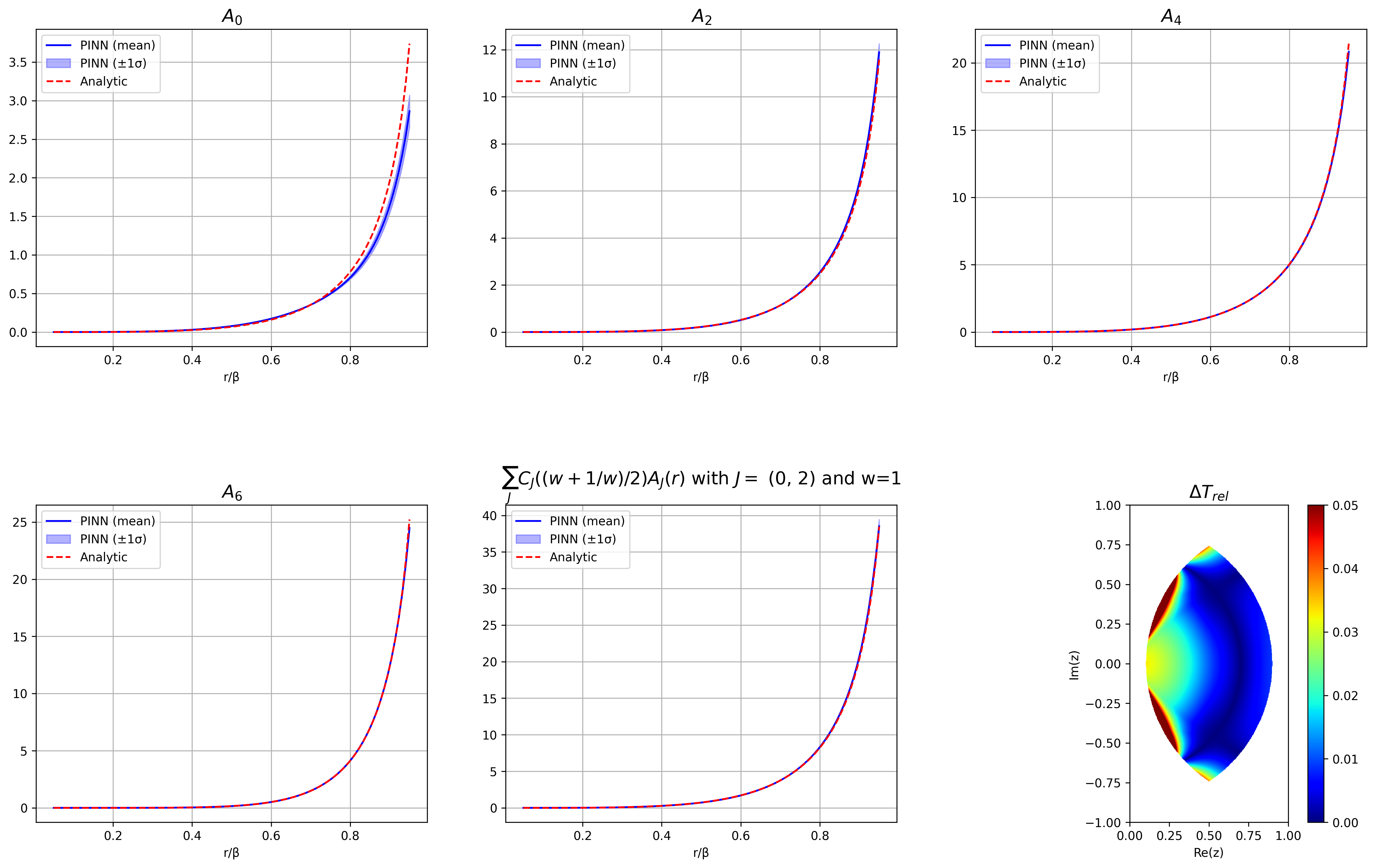}}
\vspace*{8pt}
\caption[GFF benchmark with correct analytic anchor at r=0.7]{GFF
benchmark with $d=4$, $\Delta_\phi=1.68$, $J_*=6$, one exposed
coefficient $a_{1,0}+3a_{0,2}$ and a correct analytic anchor
$\mathcal{A}_J(0.7)|_{\mathrm{GFF}}$ (case (ii) of the text). The
first four panels show the predicted tails $A_0, A_2, A_4, A_6$
(blue mean $\pm 1\sigma$ band) against the analytic GFF curves
(dashed red). The fifth panel shows the combined contribution of
$A_0$ and $A_2$ to the conformal block expansion at $w=1$; the
heatmap reports the relative difference between predicted and
analytic tails across the training region. Reproduced from
Ref.~\refcite{Niarchos:2025cdg}.}%
\label{fig:gff-anchor}
\end{figure}

The contrast between (ii) and (iii) is the core qualitative lesson
of this section. A correct anchor leaves the loss landscape with
low-loss minima clustered near the GFF solution, and the spread of
the extracted CFT datum across different initialisations is small.
An incorrect anchor still admits equally low-loss minima, but they
sit at different parameter values and the spread across
initialisations widens noticeably. This contrast, narrow ensemble
spread when the anchor is correct and broad spread when it is not,
is the empirical signature on which the stability-as-criterion
strategy of the next section relies.

\subsection{Summary of GFF lessons}

Three points carry forward to the holographic case that will be
discussed next. First, when supplied with a correct finite-$r$ anchor, the GFF
benchmark recovers all four exposed tails simultaneously at
$J_*=6$, $\Delta_\phi=1.68$ in $d=4$, in close agreement with the
analytic GFF results. Second, the approximate KMS condition is by itself
underconstrained, admitting a continuous family of low-loss
configurations, and a single correct finite-$r$ anchor is what collapses this family to a unique stable
minimum. Third, the stability of the optimisation
around the correct anchor, contrasted with the sensitivity to
incorrect anchors, offers a practical selection criterion for low-loss solutions.

%%=========================================================================
\section{Application: Holographic CFTs}\label{sec:holo}

Holographic CFTs at finite temperature provide a natural and
physically rich testing ground for the method of
Section~\ref{sec:split}. Through the AdS/CFT correspondence, at
leading order in the large-$c$ limit any classical two- (or
higher-)derivative theory of gravity in AdS furnishes, on the
boundary side, a consistent solution of the KMS condition. In
general there is an infinite family of such solutions, labelled by
CFT data dual to the couplings of the bulk gravitational action. This multi-solution structure is precisely what forced us
in Section~\ref{sec:split} to bootstrap the tails dynamically rather
than to solve a linear regression problem.

We focus throughout on a scalar operator $\phi$ with real, non-integer
scaling dimension $\Delta_\phi$ above the unitarity bound, with
vanishing thermal 1-point function $\langle \phi\rangle_\beta = 0$,
and we work in $d=4$.

\subsection{Holographic spectrum and discontinuity}

At leading order in the large $c$ limit, a scalar operator in a holographic CFT
behaves as a generalised free field coupled to the energy-momentum
sector. The operators contributing to the thermal OPE of
$\langle\phi\phi\rangle_\beta$ are collected in
Table~\ref{tab:holospectrum}: the identity, the energy-momentum tensor
$T_{\mu\nu}$, the double-twist operators $[\phi\phi]_{n,J}$, and the
multi-trace energy-momentum operators $[T^k]_J$ with $k\geq 2$ and
$0\leq 2\ell\leq 2k$. Conformal primaries with derivative insertions
on $[T^k]_J$ do not contribute at leading large-$c$; see Appendix G
of Ref.~\refcite{Niarchos:2025cdg} for the argument.

\begin{table}[t]
\tbl{Spectrum contributing to the thermal OPE of
$\langle\phi\phi\rangle_\beta$ in a holographic CFT at leading order
in large-$c$.\label{tab:holospectrum}}
{\tabcolsep14pt\begin{tabular}{@{}cccc@{}}
\toprule
Operator & $\Delta$ & Spin $J$ & Coefficient \\
\colrule
Identity & $0$ & $0$ & $a_{\boldsymbol{1}} = 1$ \\
$T_{\mu\nu}$ & $d$ & $2$ & $a_T$ \\
$[\phi\phi]_{n,J}$ & $2(\Delta_\phi + n + \ell)$ & $0\leq 2\ell$ & $a_{n,J}$ \\
$[T^k]_J$ & $d\, k$ & $0\leq 2\ell\leq 2k$ & $a^{(k)}_J$ \\
\botrule
\end{tabular}}
\end{table}

As in the GFF case, the double-twist operators $[\phi\phi]_{n,J}$
have vanishing discontinuity, so $\IT_\disc^{(\mathrm{approx})}$ is
sourced only by the energy-momentum sector (and the identity for
non-integer $\Delta_\phi$). The identity contribution is identical
to Eq.~\eqref{eq:GFFdisc}; the energy-momentum and multi-trace
pieces arise from the same crossed-channel truncation of $\disc[g]$
evaluated on the corresponding operator. Their explicit forms are
given in Ref.~\refcite{Niarchos:2025cdg}.

Two features of the holographic spectrum are worth emphasising.
First, the coefficient $a_T$ is tied to the CFT central charge via
\beq\label{eq:aT}
a_T = -\frac{2\Delta_\phi}{(d-2)(d-1)}\,
\frac{\Gamma(d/2)}{2\pi^{d/2}}\,\frac{b_T}{C_T}\,.
\eeq
Second, the lowest-twist multi-trace coefficients $a^{(k)}_{2k}$
($k\geq 2$) are \emph{universal}: a near-boundary expansion of the
bulk geometry fixes them entirely in terms of $a_T$, independently
of higher-derivative corrections \cite{Fitzpatrick:2019zqz}. The non-lowest-twist multi-trace data are no longer universal --- they depend on the choice of gravitational theory through higher orders of the near-boundary expansion --- and the double-twist coefficients $a_{n,J}$ further require the full bulk-to-boundary propagator \cite{Parisini:2023nbd}.

A natural bootstrap target is therefore the determination of the
\emph{double-twist} data from (multi-trace) energy-momentum input.
This is harder to access from gravity than the energy-momentum data
themselves, and it is the combination we focus on below.

\subsection{Fixed discontinuity from Einstein gravity}

Following Ref.~\refcite{Niarchos:2025cdg} we fix $\Delta_\phi = 1.5$
in $d=4$ and $J_* = 6$. The choice of a half-integer $\Delta_\phi$ is
made to enable direct comparison with the zero-spatial-separation
analysis of Refs.~\refcite{Buric:2025anb,Buric:2025fye}. We model
$\IT_\disc^{(\mathrm{approx})}$ by retaining all operators up to twist
$\tau_{\max} = 8$ in the energy-momentum sector,
\beq\label{eq:tauset}
\{\,T_{\mu\nu}\,\}\,\cup\,
\{[T^2]_0,[T^2]_2,[T^2]_4,[T^3]_4,[T^3]_6,[T^4]_8\}\,,
\eeq
plus the identity. This is a hard truncation, but one whose
systematic error is controlled by the kernel $\KK_{J_*}$ and
decreases with increasing $J_*$.

The corresponding coefficients predicted by Einstein gravity are
\cite{Fitzpatrick:2019zqz,Niarchos:2025cdg}
\bea\label{eq:GRdata}
&& a_{T,\mathrm{GR}} = 1.21761\,,\quad
a^{(2)}_{0,\mathrm{GR}} = -1.37668\,,\quad
a^{(2)}_{2,\mathrm{GR}} = 1.58848\,,
\nonumber\\
&& a^{(2)}_{4,\mathrm{GR}} = -4.05945\,,\quad
a^{(3)}_{4,\mathrm{GR}} = 1.77035\,,
\nonumber\\
&& a^{(3)}_{6,\mathrm{GR}} = 8.52362\,,\quad
a^{(4)}_{8,\mathrm{GR}} = -15.9641\,.
\eea
With these fixed, the approximate KMS condition \eqref{eq:kmsapprox}
becomes a search for the four tails $A_0, A_2, A_4, A_6$ and for the
exposed combination $a_{1,0}+3a_{0,2}$ accessible through the
$\LL_{\mathrm{dot}(1)}$ loss.

\subsection{ReLU-stabilised search}\label{sec:relu}

Initially running the optimisation of Section~\ref{sec:nn} with
$\LL_{\mathrm{dot}(1)}$ we encountered the same landscape of low-loss
configurations observed in the GFF analysis of
Section~\ref{sec:gff}. Motivated by the GFF observation that a finite-$r$
anchor restored stability when correct and destabilised the solution when wrong
(Section~\ref{sec:tail-aDJ}), we then implemented a soft version of that
idea through a ReLU loss centred on a reference vector $\vec A_0$ of
intermediate-radius tail values.

Explicitly, we work in the renormalised variable
$\mathcal{A}_J(r) := \mathrm{arcsinh}(A_J(r)/2)$ to compress the
several-orders-of-magnitude dynamic range of $A_J(r)$ near $r=1$, and
take as reference vector $\vec A_0 = \vec A_{\rm GFF}$, and denote
its components by $\mathcal{A}^{\rm GFF}_J(r_i)$. The ReLU term
\beq\label{eq:ReLUloss}
\LL_{\mathrm{ReLU}} = \frac{1}{J_*/2 + 1}\sum_{J}
\mathrm{ReLU}\!\Big(\,
\big|\mathcal{A}_J(r_i) - \mathcal{A}^{\rm GFF}_J(r_i)\big|
- {\tt p}\,\big|\mathcal{A}^{\rm GFF}_J(r_i)\big|
\Big)
\eeq
penalises only those configurations whose intermediate-radius tail
values deviate from the GFF reference by more than a fractional
tolerance ${\tt p}$. For small ${\tt p}$ the search is tightly
constrained near $\vec A_{\rm GFF}$ and the KMS loss is high
(because the true holographic $\vec A$ differs from
$\vec A_{\rm GFF}$); for large ${\tt p}$ the ReLU penalty is inactive
and the optimisation reverts to the unconstrained degenerate
landscape. The working hypothesis is that at an intermediate
value ${\tt p}_*$ the allowed neighbourhood just includes the true
holographic $\vec A$ and the optimisation becomes stable, i.e.
small run-to-run variance at low loss.

\subsection{Preliminary result for
\texorpdfstring{$a_{1,0}+3a_{0,2}$}{a\_\{1,0\}+3 a\_\{0,2\}}}\label{sec:holo-result}

\begin{figure}[t]
\centerline{\includegraphics[width=0.75\linewidth]{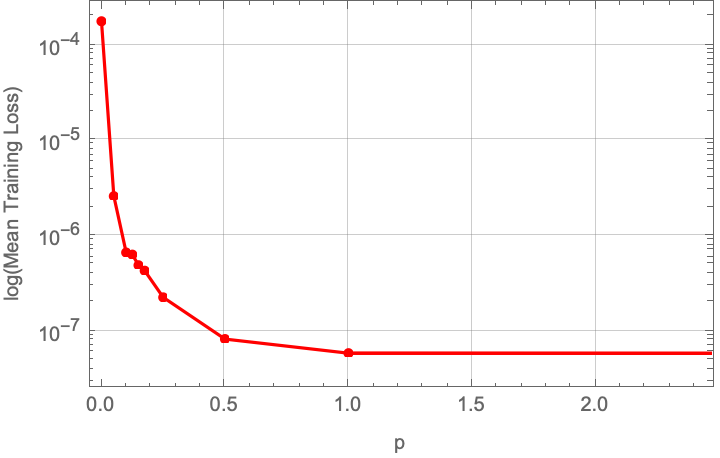}}
\vspace*{8pt}
\caption[Mean training loss as a function of the ReLU tolerance
p]{Mean training loss (log scale) of the 10 lowest-loss
configurations as a function of the ReLU tolerance ${\tt p}$, for
$1\mathrm{K}$ independent runs at each value of ${\tt p}$. A
concentration of points near ${\tt p}\simeq 0.20$ marks the regime
of minimum statistical variation. Reproduced from
Ref.~\refcite{Niarchos:2025cdg}.}\label{fig:lossvsp}
\end{figure}

Scanning ${\tt p}$ on the QMUL Apocrita cluster ($1\mathrm{K}$
independent runs, $50\mathrm{K}$ epochs per run) with
$\LL = \LL_{\mathrm{dot}(1)} + \LL_{\mathrm{ReLU}}$, we find a
pronounced stability minimum near ${\tt p}_* \simeq 0.20$ (see
Fig.~\ref{fig:lossvsp}). Extracting $a_{1,0}+3a_{0,2}$ via
\eqref{eq:aclosed} at this point gives
\beq\label{eq:holo-result}
a_{1,0}+3a_{0,2}\big|_{{\tt p}_*=0.20}
= 9.37\pm 0.44\,.
\eeq
An independent calculation of $a_{1,0}+3a_{0,2}$ employing the Pad\'e--Borel resummation of
the KMS sum rules in Ref.~\refcite{Buric:2025fye} gives $7.686$, while
the same paper's improved direct numerical PDE solve on the
AdS$_5$--Schwarzschild background returns $7.7$.

Our result \eqref{eq:holo-result}, $9.37 \pm 0.44$, is in tension with
this bulk consensus at the $\simeq\! 4\sigma$ level. We regard this as
a quantitative measure of the systematic error of the present
implementation of the ReLU-stabilised search at $J_*=6$ with the
truncated discontinuity \eqref{eq:tauset}. The
${\tt p}_*$-stability conjecture appears to be reliable to roughly
20\% but not to the percent-level for this observable. Sharpening
the prescription, through larger $J_*$, dynamical discontinuities,
or alternative monotone penalties replacing the ReLU by a quadratic
or soft-plus variant, is a natural next step.

Two sources of systematic error in \eqref{eq:holo-result} remain to
be quantified: the crossed-channel approximation of $\disc[g]$
(controllable by raising $J_*$ and $\tau_{\max}$), and the determination
of ${\tt p}_*$ itself. The key message at this stage is that our finite-$\sigma$ framework
has the structural capability to disentangle individual spin-dependent $a_{n,J}$.

%%=========================================================================
\section{Interface with Neural Spectral Bias, and Outlook}\label{sec:spectral}

The analysis of Sections~\ref{sec:gff}-\ref{sec:holo} leaves us with some interesting empirical observations. In the
GFF benchmark we found that the tail functions $A_J(r)$ cannot in
general be recovered from the approximate KMS condition on a
restricted training grid without additional input, but that a single
correct intermediate-radius \emph{anchor} $A_J(r_i)|_{\rm GFF}$
collapses the landscape of minima onto the GFF solution, while an
incorrect anchor destabilises the search and relocates it away from
the truth. In the holographic case we translated this mechanism into
a soft ReLU tolerance around a reference vector, and used
\emph{optimisation stability as a function of the tolerance} as a
signal for the true solution. At face value this is a practical
numerical recipe. However, the recent work of Refs.~\refcite{Ghosh:2026jbw,Ghosh:2026xnp} invites us to rethink the above mechanisms. This section is dedicated to
a summary of that perspective, and identifies strategies that combine different methodologies within the thermal bootstrap.

\subsection{The spectral-bias perspective}\label{sec:sb-recap}

Refs.~\refcite{Ghosh:2026jbw,Ghosh:2026xnp} consider single-variable
CFT correlators on $z\in(0,1)$ in two kinematic settings:
line-restricted four-point functions on the diagonal $z=\bar z$, and
thermal two-point functions at zero spatial separation on
$S^1_\beta \times \IR^{d-1}$. In each case the
correlator is parametrised as
\beq\label{eq:GKNSsetup}
\mathcal{G}(z) = L(z) + H(z)\,,\qquad
H(z) = z^\delta\, \mathrm{NN}_{\boldsymbol{\theta}}(z)\,,
\eeq
where $L(z)$ is a known leading-behaviour piece dominant as $z\to 0$
and $H(z)$ is represented by a light feed-forward MLP with smooth
activations (tanh or GELU) $\delta$ is known input on the leading small-$z$ behaviour of the function $H(z)$. The networks are optimised on a
context-appropriate crossing or KMS equation. In the
line-restricted case, we impose the crossing equation
\beq\label{eq:GKNScross}
\mathcal{G}(z) = \left(\frac{z}{1-z}\right)^{2\Delta_\phi}\mathcal{G}(1-z)\,,
\eeq
together with a single anchor $\mathcal{G}(z_0) = H_0$ condition at one (arbitrarily chosen) point $z_0\in(0,1)$. Trained in this way, NNs with only a few
thousand parameters reconstruct the full functional form of
$\mathcal{G}(z)$ on the interval $z\in(0,1)$ to percent-level accuracy
across a broad panel of examples: AdS$_2$ contact and one-loop Witten
diagrams, generalised free fields, unitary and non-unitary 2d minimal
models, the 3d Ising CFT, 4d half-BPS correlators in $\NN=4$ SYM, and
the 3d Ising thermal $\langle\sigma\sigma\rangle_\beta$ and
$\langle\epsilon\epsilon\rangle_\beta$ correlators. The same construction extends to the full $(z,\bar z)$ plane via crossing on concentric circles around the crossing-symmetric point $z=\bar z = \tfrac12$, seeded by the line-restricted reconstruction.  

The robustness of this approach across theories and dimensions is
striking because the crossing equation alone vastly underdetermines
the correlator: for any solution $H_1(z)$, any function
$H_2(z) = H_1(z) + (1-z)^{-2\Delta_\phi} f(z)$ with $f(z) = f(1-z)$
and $f(z_0)=0$ is also a solution. Why does gradient-based training, run on
an MLP with a small number of parameters, prefer the \emph{physical}
solution among this infinite-dimensional family?

The answer offered in Refs.~\refcite{Ghosh:2026jbw,Ghosh:2026xnp} is
\emph{spectral bias}, a well-documented phenomenon in deep learning:
gradient-descent training of an MLP with smooth activations learns
low-frequency components of a target function before high-frequency
ones, and, in the infinite-width limit, where the training is
governed by the Neural Tangent Kernel (NTK), implicitly minimises
a Reproducing Kernel Hilbert Space (RKHS) norm that heavily
penalises sharp, high-curvature, oscillatory behaviour \cite{Jacot:2018ivh,John_Xu_2020,rahaman2019spectralbiasneuralnetworks,bietti2019inductivebiasneuraltangent}. The
empirical observation of Refs.~\refcite{Ghosh:2026jbw,Ghosh:2026xnp}
is that \emph{physical CFT correlators are smooth functions} in a
precise quantitative sense (small fractional Sobolev semi-norms,
fast-decaying Chebyshev coefficients, low curvature functionals), so
that the implicit smoothness prior of gradient descent selects them
from the crossing-symmetric solution space. The crossing equation
does the rest of the work, and the single anchor is enough to fix the residual ambiguity.

\subsection{Towards improved strategies and open problems}\label{sec:sb-reading}

Despite their common philosophy, our deep conformal bootstrap framework and that of
Refs.~\refcite{Ghosh:2026jbw,Ghosh:2026xnp} are initially targeted at different
kinematical regimes and different pieces of CFT data, and the two are
naturally complementary.
Our framework receives specific theory input through dispersion relations and resolves the full two-variable structure of $g(rw, rw^{-1})$ by emphasising the use of spin-dependent tail functions. On the other hand, the anchored neural conformal bootstrap approach of Refs.~\refcite{Ghosh:2026jbw,Ghosh:2026xnp} is more minimalistic and reconstructs the correlator along sequences of one-dimensional sections.   
In that light, we find the following directions worth pursuing further:

\begin{itemlist}
\item It would be useful to better understand the potential relation between the numerical stability of the NN optimisation and the choice of the anchor points in the spin-dependent tails observed in our framework, as well as the interplay with the dispersion relation input, from the perspective of the NN spectral bias. It would also be interesting to explore to what degree certain lessons about correlator smoothness distilled in  Refs.~\refcite{Ghosh:2026jbw,Ghosh:2026xnp} can carry over to the setup of the deep conformal bootstrap summarised in this paper.

\item Since the two aforementioned approaches are complementary, one might expect to obtain more powerful results by combining them into a single framework that employs the simultaneous use of different representations of the same thermal correlators. In addition, one could also input results obtained within other numerical and analytic thermal bootstrap approaches developed recently \cite{Marchetto:2023xap,Barrat:2025wbi,Barrat:2025nvu,Buric:2025fye}. In this context, we hope that the incorporation of a \emph{dynamical discontinuity} contribution to the dispersion part can be implemented more efficiently.

\end{itemlist}

Within such a hybrid pipeline, one could further explore the thermal physics of several different physical systems. Of particular interest are thermal correlators of holographic CFTs, at and beyond the SUGRA regime, and physically relevant CFTs like the $O(N)$ vector models in 3d. 

More generally, it is worth exploring if similar methods --- combining input from different sources, including dispersion relations, smoothness properties, as well as analytic and numerical conformal bootstrap results --- can be extended beyond finite-temperature two-point functions. CFTs with defects or
boundaries and higher-point bootstrap, where positivity is also absent, stand out as natural targets.

%%=========================================================================
\section*{Acknowledgements}

It is a pleasure to thank our collaborators A.~Stratoudakis and
M.~Woolley for collaboration in the work that underpins this
proceedings contribution. VN would also like to thank Kausik Ghosh, Sidhaarth Kumar and Andreas Stergiou for collaboration on Refs.~\refcite{Ghosh:2026jbw,Ghosh:2026xnp} and related topics. CP acknowledges the organisers of the
``Athens Workshop in Theoretical Physics: 10th Anniversary" for the
kind invitation to contribute to its proceedings. The work of CP was
partially supported by the Science and Technology Facilities Council
(STFC) Consolidated Grant ST/X00063X/1 ``Amplitudes, Strings \&
Duality''. This research utilised Queen Mary's Apocrita HPC
facility, supported by QMUL Research-IT:
\href{http://doi.org/10.5281/zenodo.438045}{http://doi.org/10.5281/zenodo.438045}.

%%=========================================================================

\bibliographystyle{ws-ijmpd}
\bibliography{finiteT}

\end{document}